# Fluid Motion Physics:
# The Total Head Vector of the Real Fluid Stream

## S. L. Arsenjev[1]

*Physical-Technical Group*
*Dobroljubova Street, 2, 29, Pavlograd Town, 51400 Ukraine*

The overcoming of a mechanics problem on origin of secondary jet flows, on dynamics of its development and interaction with the main stream of the viscous fluid is reached by means of elucidation of the energy distribution in the stream at its interaction with the flow system walls and a body surface. In contrast to a traditional conception on a total head of a fluid stream in a pipe as a sum of a velocity head and a static pressure, it is ascertained that this sum is only longitudinal component of the stream total head vector. A new conception is stated and the expressions for quantitative evaluation of radial and tangential components of the stream total head vector and for angle of its deflection from the stream axis are produced. The new approach has allowed rationally to explain the causes of origin of the secondary jet flows and dynamics of its interaction with the main stream in various cases. Thereby with the taking into account of previous articles of the author the necessary theoretical preconditions has been created for a working out of the modern physically valuable apparatus for mathematical modeling of the real – viscous – fluid motion in any problems of hydromechanics and gas dynamics.

## Introduction

A beginning of intensive development of hydrodynamics is bound with November 1732 when H. Pitot has invented a device allowing simultaneously to measure both static pressure and a sum of the static pressure and velocity head in the real liquid (water) stream. Various modifications of this device are wide applied in modern researches of the liquid and gas motion.

After 6 years from Pitot remarkable invention D. Bernoulli in his monograph "Hydrodynamics" has adduced an analogy in accordance with which the static pressure and the velocity head of the liquid stream in a pipe were respectively compared with potential and kinetic energy of a free falling body, i.e. without the taking into account of the resistance forces to the motion. On the face of it, a correctness of such analogy is corroborated by measurements by means of Pitot device. Difference is only that Pitot device allows to measure the liquid stream parameters with the taking into account of the stream interaction with the pipe wall, i.e. with the taking into account of the forces of the liquid internal friction, while Bernoulli analogy surmises an absence of interaction of the liquid stream with the pipe wall, i.e. it is surmised a fluid, not possessed by viscosity. Use of Pitot device has allowed to determine all three initial components of the stream energy: static pressure under action of the gravitation force, static pressure under action of the viscosity force and the velocity head under any orientation of the pipe. Side by side with it Pitot device has allowed to determine the static pressure distribution under action of hydraulic friction along the pipe and the velocity profile in any cross-section of the pipe.

Thus, just Pitot device has allowed to create physically adequate (but not speculative) preconditions for fruitful development of hydro- and aeromechanics and gas dynamics.

However subsequent development had led to a division of a science on the fluid motion into theoretical (speculative) and applied (empirical) dynamics of fluid. As to the first of these scientific directions, O. Tietjens [1] was writing: "The flow phenomena it failed to make accessible to a mathematical research till now" and he was adding: "This field is so fully investigated, especially mathematically, that in essential it can be considered as a finished one". S. Goldstein [2] was writing: "Mathematical calculations in the boundary-layer theory are based on the pressure distribution, given beforehand. The boundary-layer theory cannot be used for its determination". After 50 years, Yu. V. Lapin and M. Kh. Strelets [3] were writing with disappointment: "…the simple qualitative conceptions about nature of the studied phenomena … always were by a good basis for creation of its models". The second of these scientific directions, based on experiment, allows in the second half of the $18^{th}$ century to find the correct form of the outflow equation – the so-called Torricelli formula, in the middle of the $19^{th}$ century to find the correct expression for the head loss because of the stream friction along the pipe – Darcy - Weissbach equation and, at last, at the close of the $19^{th}$ century to give the correct form of the energy conservation law for the liquid stream in the pipe with the taking into account of a friction – the so-called Bernoulli equation.

---
[1] Phone: (+3805632) 40596 (home, Rus.)
E-mail: usp777@ukr.net, ksenia5766@mail.ru

In the $20^{th}$ century the applied hydrodynamics enriches oneself by the new methods and instruments for experimental research of the flows and at the same time both directions of hydro- and gas dynamics enrich themselves by the high-powered computer engineering techniques. However a presence of the developed instrumental basis for experimental and theoretical researches in the field of hydromechanics was found insufficient for elucidation of nature of such phenomena as, for example, the secondary flows. These phenomena are highly desirable in the plants for a mixing of fluids and these phenomena are highly not desirable in a motion of the ground, air and water transport.

The object of the given article is just in elucidation of mechanics of the secondary flows mainly in the flow systems and its elements.

**Approach**

For solution of the raised problem it is expediently to divide the secondary flows by way of its formation into the following types:

- the secondary flows, formed by the wall or body geometry;
- the secondary flows, formed by friction against a wall.

Taking into account that both experimental and theoretical methods and instruments not allow till now to find the causes of origin of the secondary flows, formed by friction, the author supposes that it is necessary to find such experiment, which one will distinctly indicate a cause of its origin.

**Solution**

As the crucial experiment, allowing immediately – without a measuring instrument – to observe an action of a friction upon the secondary flow origin, the author considers the simplest experiment accessible for realization in the house conditions.

Hold in one your hand horizontally over bath the plate $\sim 20 \times 30 cm$ of transparent unbreakable material (resin glass). Direct by the other your hand a water jet $\sim 1 cm$ diameter – by means of the flexible hose, connected with a tap, – along the upper surface of the plate. Regulating the water flow, it can be made that the water jet, passing the greater part of the plate length, comes to a short stop with formation of the so-called hydraulic jump. Water falls freely out of the jump over the plate edge into the bath. On a surface of the plate one can see that the originally circular water jet spreads into both sides from its flow direction in the time of its motion along the plate. One may think that the jet becomes flat under action of its own weight. Then direct this jet along the lower surface of the same horizontal plate. One can see that the jet, moving along the lower surface of the plate, spreads and becomes flat the same as on the upper surface of this plate; then one the same comes to a short stop and falls plumb out of hydraulic jump into the bath.

Diagrams of this experiment are presented in figure.

In this experiment the water jet, moving along the plate, undergoes action of four types of forces: a surface tension, gravity, atmospheric pressure and the liquid friction against the plate surface.

The surface tension forces strive to prevent the spreading and a flattening of the jet.

Gravity is the same all over the plate and one is directed downward.

Atmospheric pressure is the same around the plate and one not influences the jet form. At the same time atmospheric pressure presses the jet to the plate surfaces. Therefore the jet "adheres" to the plate lower surface. [This "adhering" of liquid to a wall is stipulated by the continuous contact – without gaps – what is characteristically for the fluid – wall contact. In contrast to it the unlubricated – dry - contact of the solid bodies is a discrete (point) contact because of a roughness of its surfaces. Therefore an air streak reduces an action of the adhering to insignificant quantity in the contact zone of the solid bodies].

An interaction of the liquid stream (including the liquid jet) with the wall is characterized by the well-known features. In the first, a contact of the liquid stream with the solid body surface is accompanied by a physisorption of molecules of the fluid, which one forms a wall boundary layer by 50 – 500 diameters of the molecules. Viscosity and density of the fluid in this layer is significantly increased towards a solid surface. Therefore the most part of the molecules in this layer remains immovable in the time of a motion of the liquid stream. In particular, in the flow laminar regime a resistance to the fluid motion is determined by internal friction of the fluid but not a roughness of the solid surface. In the second, the fluid stream motion is accompanied by formation of the shear layer over the physisorption layer. This is a layer of the most intensive shear. The first signal of it was given by P.L.G. Du Buat (1786), i.e. almost 120 years before the boundary-layer theory, advanced by L. Prandtl. Thickness of this layer can be determined, in a quantity order, by the well-known expression

$$\delta = k\sqrt{\nu \cdot l / v_m},$$

where $k = 3.64$ is a dimensionless coefficient, $\nu \cong 10^{-6}\ m^2/s$ is a water kinematics viscosity, $l = 0.2m$ is the jet length on the plate, $v_m$ is a mean velocity in the jet cross-section.

In the above-described house experiment a thickness of the intensive shear layer in the jet end not exceeds approximately

$$\delta_l = 3.64\sqrt{10^{-6} \cdot 0.2/2} = 1.15mm.$$

A presence of the most intensive shear layer one can see, pouring out not great quantity of water out of usual bucket or even out of a cup.

Now compare the considered motion of the liquid free jet along horizontal plate with a motion of the liquid stream in horizontal pipe. It is well-known that interaction of the liquid stream with pipe wall is accompanied by an increase of static pressure in a direction from the pipe outlet to the pipe inlet. The static pressure drop is determined by Darcy - Weissbach equation

$$h_{st} = \lambda \frac{l}{d} \cdot \frac{1}{g} \cdot \frac{v_m^2}{2} \equiv \frac{\lambda \cdot \bar{l}}{g} \cdot \frac{v_m^2}{2},$$

which one is transformed into Torricelli formula when $\lambda \cdot \bar{l} = 1$.

Now suppose that the liquid jet moves along the plate also under a constant velocity, i.e. the jet changes only its cross-section form but one not changes a quantity of its cross-section area. Then one can conclude that a flattening of the jet, moving along the plate, has the same nature with static pressure of the stream in the pipe. Difference between them is only that the pipe wall restricts a possibility of a spreading of the liquid stream and thereby one leads to an accumulation of static pressure in the liquid stream. Absence of lateral restriction of the liquid free jet, moving along the plate, allows its lateral spreading, what is equivalently – with a point of view of the energy conservation law – to a static pressure of the stream in a pipe. A cause of static pressure in the pipe stream and of a flattening of the liquid free jet on a plate is only one – internal friction in a fluid stream, moving along the solid surface, i.e. viscosity of a fluid.

Continuing a considering of the experiment with the liquid free jet on the plate, introduce a system of the orthogonal coordinates with its reference point O, coinciding with a beginning of the jet on the plate, at that axis OX is directed along the jet longitudinal symmetry axis, axis OY is parallel and axis OZ is perpendicularly to the plate surface; at the same time XOZ is the jet longitudinal symmetry flatness. Then, according to Darcy - Weissbach, one can write

$$h_{st} \equiv h_y = \frac{\lambda \cdot \bar{l}}{g} \cdot \frac{v_x^2}{2} \qquad (1)$$

from where

$$v_y = \sqrt{2gh_y} = \sqrt{2g \frac{\lambda \cdot l}{g} \cdot \frac{v_x^2}{2}} = v_x \sqrt{\lambda \cdot \bar{l}}. \qquad (2)$$

Ratio

$$v_y / v_x = \sqrt{\lambda \cdot \bar{l}} = \tan \alpha_{yv} \qquad (3)$$

determines kinematically an angle of deflection of the jet lateral boundaries in both sides from its longitudinal symmetry flatness XOZ. Condition $\bar{l} = 0$ signifies a beginning of the jet where $\alpha_{yv}^0 = \arctan \sqrt{\lambda}$ is maximum. When $Re \to 0$, then $\lambda \to 0.101$ and $\alpha_{yv}^0 \to 17.63^0$; when $Re \to 2.5 \cdot 10^6$, then $\lambda \to 0.01$ and $\alpha_{yv}^0 \to 5.72^0$. In the other words, the more initial velocity of the jet the lesser angle of deflection of the jet lateral boundaries in its beginning from its longitudinal symmetry flatness.

Thus the experiment with the liquid free jet, moving along horizontal plate, leads to a necessity to distinguish a velocity head in longitudinal direction (along the abscissa axis) $p_x = \rho \cdot v_x^2 / 2$, and a velocity head in transversal direction (along the ordinate axis in both sides from flatness XOZ) $p_y = \rho \cdot v_y^2 / 2 = \lambda \cdot \bar{l} \cdot \rho \cdot v_x^2 / 2$ as the vector quantities. At that the jet total head vector is determined by geometrical sum of these components

$$p_\Sigma = \sqrt{p_x^2 + p_y^2} = \rho \cdot \frac{v_x^2}{2} \sqrt{1 + \left(\lambda \cdot \bar{l}\right)^2}, \qquad (4)$$

and angle of its deflection in both sides from the jet longitudinal symmetry flatness XOZ is determined respectively by expression

$$\alpha_{yp} = \arctan p_y / p_x = \arctan v_y^2 / v_x^2 = \arctan \lambda \cdot \bar{l} \qquad (5)$$

with its maximum quantity in a beginning of the jet

$$\alpha_{yp}^{max} = \arctan \lambda. \qquad (6)$$

Observing the water jet motion along the horizontal plate, one can see the wave regular structure, formed by longitudinal and transversal components of the stream total head vector.

A motion of the liquid stream in horizontal cylindrical pipe is, to a certain degree, similar to a motion of the liquid jet along the horizontal plate as a pipe is the right-angled plate, rolling up with a constant radius up to a contiguity of its opposite sides. A closed perimeter of the pipe cross-section stipulates an origin of additional – radial – component of the stream total head vector. For explanation, introduce a system of orthogonal coordinates with its reference point O in some point on the pipe longitudinal axis; at that axis OX coincides with the pipe longitudinal axis, axis OZ is directed along the pipe radius and axis OY is in the pipe cross-section flatness. Components of the stream total head vector:

- dynamic component
$$p_{dx} = \rho \cdot \frac{v_x^2}{2},$$

- static components:
$$p_{stx} = p_{sty} = p_{stz} = \lambda \cdot \bar{l} \cdot \rho \frac{v_x^2}{2}. \tag{7}$$

The stream total head vector in horizontal cylindrical pipe is determined by expression

$$p_{\Sigma} = \sqrt{(p_{dx} + p_{stx})^2 + p_{sty}^2 + p_{stz}^2} = \rho \frac{v_x^2}{2} \sqrt{(1 + \lambda \cdot \bar{l})^2 + 2(\lambda \cdot \bar{l})^2}. \tag{8}$$

The closed perimeter and central symmetry of the pipe cross-section exclude possibility of transversal spreading of the liquid stream in the pipe in contrast to a motion of the liquid free jet along the plate. In the pipe a transversal velocity head of stream can be realized only in the kind of the hoop – tangential – compressive stress, provoking equal to it by quantity the radial compressive stress, called in hydromechanics by static pressure. In the result an arithmetical sum of the velocity head and static pressure in the same direction determines the longitudinal component of the stream total head vector in the pipe

$$p_{\Sigma x} = \rho \cdot \frac{v_x^2}{2} + \lambda \cdot \bar{l} \cdot \rho \frac{v_x^2}{2} = \rho \frac{v_x^2}{2} (1 + \lambda \cdot \bar{l}). \tag{9}$$

Radial component of the stream total head vector in the pipe is a geometrical – vector – sum of the longitudinal component and static pressure in radial direction

$$p_{\Sigma xz} = \sqrt{p_{\Sigma x}^2 + p_{stz}^2} = \rho \frac{v_x^2}{2} \sqrt{(1 + \lambda \cdot \bar{l})^2 + (\lambda \cdot \bar{l})^2}. \tag{10}$$

Correspondingly an angle of deflection of radial component of the stream total head vector from the stream axis in the pipe is determined by expression

$$\alpha_{\Sigma xz} = \arctan \frac{\lambda \cdot \bar{l}}{1 + \lambda \cdot \bar{l}}. \tag{11}$$

It is necessary to explain that the expressions (10, 11) determine the radial component quantity and its deflection angle along the pipe by means of its relative (caliber) length. At that it is necessary to measure the pipe length from its outlet end. Thus at an outflow of the liquid stream out of the pipe $\bar{l} = 0$ and expression (11) takes the kind

$$\alpha_{\Sigma xz} = \arctan \frac{0}{1 + 0} \rightarrow 0, \tag{12}$$

testifying to that the liquid free jet, flowing out of the pipe, will be having the pipe cross-section. At the same time in the inlet end of the sufficiently long pipe $\bar{l} \gg 1,$ therefore expression (11) takes the kind

$$\alpha_{\Sigma xz} = \arctan \frac{\bar{l}}{\bar{l}} \rightarrow 45^0. \tag{13}$$

One of substantial properties of radial component of the stream total head vector is that this component is directed to a surface of a contact of the stream with a wall. When this component is directed away from the contact surface then it is going on a separation of the stream from the wall. Tangential component of the stream total head vector in the pipe is geometrical sum of its longitudinal component – according to the expression (9) – and static pressure in tangential direction. Formally the expression for this component has the kind similar to the expression (10)

$$p_{\Sigma xy} = \sqrt{p_{\Sigma x}^2 + p_{sty}^2} = \rho \cdot \frac{v_x^2}{2} \sqrt{\left(1 + \lambda \cdot \bar{l}\right)^2 + \left(\lambda \cdot \bar{l}\right)^2}, \qquad (14)$$

however, in contrast to radial components this component is directed symmetrically in both sides from radial flatness XOZ under angle

$$\alpha_{\Sigma xy} = \pm \arctan \frac{\lambda \cdot \bar{l}}{1 + \lambda \cdot \bar{l}}. \qquad (15)$$

The expression (15) has the same features with expressions (12, 13).

An action of the radial component can be physically interpreted so: if the pipe will be exerting a resistance to the fluid stream motion and at the same time the pipe will not be having a hoop strength and rigidity, then the stream will be spreading in the kind of a cone with angle of deflection of its generatrix from its axis equal to $\alpha_{\Sigma xz}$. In a real – strong and rigid – cylindrical pipe the conical spreading is impossible, therefore the stream total head vector determines only a possibility of radial and tangential spreading of the stream in the pipe and thereby one determines an ability of the stream to a formation of the secondary flows under conditions of deflection of the pipe axis from a straight line, of a change of the pipe cross-section area, of deflection of the pipe cross-section form from circular one.

Feature of tangential component of the stream total head vector is that this component turns from potential state of static pressure into active state of a velocity head under conditions of the flatness symmetry as it is going on, for example, in the time of a motion of the liquid free jet along the plate. Under conditions of axial symmetry of a pipe with its circular cross-section the tangential component remains in its potential state of static pressure.

Till now in this section the author supposed, for a simplification, the stream velocity as a constant quantity all over the stream cross-section. Conformably to the liquid stream in a pipe with the well-known velocity profile – parabolic in a case of the laminar flow regime and closed to logarithmic in a case of the turbulent flow regime – it is necessary the chosen cross-section of the stream to divide into elementary concentric layers and to determine a quantity and the orientation angle of components of the stream total head vector for every such layer. For example, the sufficiently sharp contraction of the pipe cross-section signifies a radial resistance to the liquid stream motion. A determination of a quantity and an incline angle of radial component of the stream total head vector immediately in front of the contraction zone allow to ascertain not only a possibility but also necessity of origin of the secondary return flow in this zone.

With the taking into account of a change of the stream motion velocity in the chosen cross-section of the pipe an expression for radial component of the stream total head vector in elementary annular layer has the kind

$$p_{\Sigma xzi} = \sqrt{p_{\Sigma xi}^2 + p_{stl}^2} = \sqrt{\left(p_{stl} + \rho \frac{v_{xi}^2}{2}\right)^2 + p_{stl}^2}, \qquad (16)$$

and expression for the deflection angle of the radial component has the kind

$$\alpha_{\Sigma xzi} = \arctan p_{stl} \Big/ \left(p_{stl} + \rho \cdot v_{xi}^2 / 2\right). \qquad (17)$$

In expressions (16, 17) subscript $l$ signifies a quantity of the stream static pressure, determined by Darcy – Weissbach equation. Subscript $i$ signifies the elementary annular layer number within the limits of the pipe cross-section radius before the pipe contraction. Expression (17) testifies to that a quantity of the deflection angle of the radial component is the least, but one is already greater naught, by the stream axis and this angle equal to $45^0$ by the pipe wall.

Expressions for a quantity and the deflection angle of the tangential component from the stream axis are quite analogically to the expressions (16, 17).

With the taking into account of the velocity profile in the pipe cross-section the expressions (9, 16 and 17) can also be represented in the kind:

- longitudinal component

$$p_{long} = \rho \, \frac{v_x^2(r)}{2} + \lambda \cdot \bar{l} \cdot \rho \, \frac{v_{xm}^2}{2} = \lambda \cdot \bar{l} \cdot \rho \, \frac{v_{xm}^2}{2} \{ \frac{1}{\lambda \cdot \bar{l}} \left[ \frac{v_x(r)}{v_{xm}} \right]^2 + 1 \}, \qquad (18)$$

- radial and tangential components

$$p_{rad} = p_{\tan} = \lambda \cdot \bar{l} \cdot \rho \, \frac{v_{xm}^2}{2} \sqrt{ \{ \frac{1}{\lambda \cdot \bar{l}} \left[ \frac{v_x(r)}{v_{xm}} \right]^2 + 1 \}^2 + 1 }. \qquad (19)$$

- the deflection angle of radial component from the stream axis and of tangential component in both sides from radial flatness

$$\alpha_{rad} = \alpha_{\tan} = \arctan \frac{1}{\frac{1}{\lambda \cdot \bar{l}} \left[ \frac{v_x(r)}{v_{xm}} \right]^2 + 1}. \qquad (20)$$

Conformably to the liquid stream in a straight pipe of a round cross-section a tangential component of the stream total head vector is retained in the kind of static pressure, directed symmetrically in both sides from the stream radial flatness under angle $45^0$ by the pipe wall. In that way a trajectory of the tangential component action forms two cylindrical – left-hand and right-hand – helices. Intensity of static pressure along the helices corresponds to a static pressure along the pipe: from naught by the outlet end of the pipe to a maximum by the inlet end of the pipe. An incline of the helices is decreased from the pipe wall to the stream axis in accordance with expressions (17, 20). Thus a motion of a real fluid in a pipe forms in its stream a kinetostatics state, answering to Pascal law in the stream cross-section flatness and at the same time corresponding to the stressed state of a steel bar of cylindrical form under action of own weight along its length [4, 5].

The common character of the energy distribution in the real fluid stream allows to write expressions for the components of the gas stream total head vector. In particular, expression for the longitudinal component has the kind

$$p_{\Sigma x} = \rho \, \frac{v_x^2}{2} + p_{stx} = \rho \, \frac{v_x^2}{2} + \rho \, \frac{c^2}{k} = \rho \, \frac{c^2}{k} \left( \frac{k}{2} M^2 + 1 \right), \qquad (21)$$

where $v_x$ is a mean velocity in the stream cross-section, $c$ is a sound wave velocity, $k$ is a ratio of the specific heats of the gas, $M$ is Mach number.

Expression for the radial component has the kind

$$p_{\Sigma xz} = \sqrt{\left[\rho \frac{c^2}{k}\left(\frac{k}{2}M^2+1\right)\right]^2 + \left(\rho \frac{c^2}{k}\right)^2} = \rho \frac{c^2}{k}\sqrt{\left(\frac{k}{2}M^2+1\right)^2+1} \ . \qquad (22)$$

Angle of deflection of the radial component from the stream axis

$$\alpha_{\Sigma xz} = \arctan \frac{1}{\frac{k}{2}M^2+1} \ . \qquad (23)$$

With the taking into account of the velocity profile in the pipe cross-section the expressions (21 - 23) can also be represented in the kind:
- longitudinal component

$$p_{long} = \rho \frac{v_x^2(r)}{2} + \rho \frac{c^2}{k} = \rho \frac{c^2}{k}\left[\frac{k}{2}M^2(r)+1\right] \ , \qquad (24)$$

- radial and tangential components

$$p_{rad} = p_{\tan} = \rho \frac{c^2}{k}\sqrt{\left[\frac{k}{2}M^2(r)+1\right]^2+1} \ , \qquad (25)$$

- the deflection angle of radial component from the stream axis and of tangential component in both sides from radial flatness

$$\alpha_{rad} = \alpha_{\tan} = \arctan \frac{1}{\frac{k}{2}M^2(r)+1} \ . \qquad (26)$$

Solving any problem on a motion mechanics of the real fluid, it is necessary to combine the stream total head vector with a possible action of centrifugal forces, stipulated by a form of the solid surface, and – in a case of necessity – to take into account the external (atmospheric or hydrostatic) pressure action, the gravitation forces, the surface tension and the same time to take into account its vector character, determined by geometry of the interaction surface of the fluid stream with the wall.

**Discussion of results**
Now, moving successively from the simple to the complex, apply a notion of the stream total head vector (STHV) to an elucidation of the causes of origin of the secondary flows in different cases. Imagine to yourselves a liquid stream in straight flume of rectangular cross-section. Similar case was considered by J. Thomson (1878) [2] and he has adduced his explanation to the causes of origin of the secondary flow in this and similar to it cases.
A liquid motion in this flume the same way as in the case of the liquid free jet motion along the plate possesses two vertical flatness of symmetry, passing along and across the liquid stream. At that it should be distinguished the wide flume with its breadth sufficiently greater then its depth and the narrow flume with its breadth commensurable with its depth.
Consider in the beginning a flow in the wide flume. A presence of the flat surfaces of its bottom and lateral walls, of a free surface of the liquid stream between its lateral walls in combination with a longitudinal and transversal symmetry of the stream allows to use for the given case the results of consideration of the liquid free jet motion along the plate. As a matter of fact, the liquid stream motion in the flume is distinguished only by a presence of the walls, restricting the lateral spreading of the stream. The same way as in the case of the jet motion along the plate is accompanied by the transversal spreading of the most intensive shear layer by the flume bottom into both sides of its longitudinal flatness of symmetry. Here the deflection angle of transversal component of STHV from the flow axis is determined by expression (6). Under the same angle the transversal flow runs against the flume lateral walls, then one moves along and upwards the walls and further one moves under the same angle on the stream upper surface to its longitudinal flatness of symmetry. In that way by every of two lateral walls of the flume it is arisen the secondary flow with its trajectory in the kind of a multistart cylindrical helix, formed by the elementary vortex threads. A pitch of the helices is determined by a maximum quantity of the deflection angle of transversal component of STHV. Number of the elementary vortex threads in one pitch of the helices is determined by effective thickness of the most intensive shear layer by the flume bottom. Diameter of these cylindrical helices is determined by the stream depth in the flume. If the flume breadth is commensurable with the doubled diameter of the secondary flow helices, then the secondary flows come running together on the free surface of the stream and here ones fill, in partially, a decrease of longitudinal component of STHV because of the stream friction against the flume bottom.

L. Prandtl in his book [6] has adduced the results of visualization of a flow in the straight pipe with its cross-section in the kind of equilateral triangle, obtained by J. Nikuradze (1930). The liquid stream, moving forwardly, forms in the pipe cross-section the secondary pair-spiral flows by every apex of the triangle with the sum flows from a middle of the triangle sides to its apices.

With geometrical point of view the equilateral triangle possesses a rotary central symmetry of the third order. Therefore this triangle can be formed by a successive turning onto $120^0$ of one of its elements – side or corner – around its centre. Choice of the corner as an element, forming this triangle, allows to present it as the flume cross-section. It is not difficult to understand from previous example that the forward flow in a flume with its cross-section in the kind of a corner with its angle equal to $60^0$ forms the pair-spiral secondary flow with its sum direction to its vertex. In the result the flow in the straight pipe with its cross-section in the kind of equilateral triangle can be presented as a sum of the flows in the same three flumes, forming a cross-section of such pipe.

S. Goldstein in the above mentioned book [2] has adduced the diagram and short description of an elegant experiment, conducted by G. I. Taylor (1929): the water laminar stream moves in the glass pipe of a round cross-section; this straight pipe has a section of the local smooth crook with its maximum deflection from the pipe axis some lesser then the pipe diameter; the thin jet of a colored liquid moves through the crook practically without deflection from a straight line.

To elucidate such strange, on the face of it, conduct of the colored jet it should be paid attention to that the mentioned local crook, in contrast to an axial symmetry of the straight pipe, possesses the flatness-mirror symmetry, determined by its curvature flatness. To the point, in Taylor experiment the colored jet moves in the curvature flatness of the local crook and one has been displaced from the pipe cross-section centre. The local crook is formed by four same turns: two to the left and two to the right with a return to the straight pipe axis. Before the local crook the pipe axial symmetry stipulates potential state of the radial and tangential components of STHV. The flatness-mirror symmetry of the first turn of the local crook stipulates transition of tangential component of STHV into its kinetic state in the kind of a velocity head. As a result this component of STHV transforms a primary simple forward flow into the pair-spiral flow with a spreading of the water wall layer into both sides from the symmetry flatness and along the pipe cross-section contour in direction to a curvature centre of the first turn. The total pair-spiral flow is going on along the symmetry flatness and in direction away from a curvature centre of the first turn of the pipe. This total flow displaces the colored jet in direction away from the curvature centre of the first turn of the pipe. At that a proportionality of intensity of the pair-spiral flow to a curvature and a turn angle of this crook section ensures a straightforwardness of the colored jet trajectory. The second turn of the local crook is distinguished from its first turn only by a reverse curvature. Accordingly to it the reverse pair-spiral flow in the second turn reduces smoothly to a naught the pair-spiral flow out of the first turn of the crook. As a result the colored jet to the end of the second turn is displaced into a diametrically opposite point in the pipe crook cross-section relative its initial position in the straight pipe cross-section before the crook, and the colored jet is continuing its motion along a straight line. In two subsequent turns of the local crook the above-described process is repeating and one ensures further rectilinear motion of the colored jet and its smooth return into an initial position in the straight pipe cross-section after its local crook.

As regards centrifugal forces, arising in a liquid stream in the curvilinear pipe, their action, as it is known, comes only to a creation of hydrostatic pressure proportional to an angular velocity and a rotation radius. Therefore a cause of origin of the secondary pair-spiral flow in the curvilinear pipe it should be found only on basis of the contact interaction of a real – viscous – fluid with a solid surface: in the given case – in the interaction by friction of the liquid stream with a wall of the pipe, possessing deflections both from axial symmetry and (or) from a constancy of the pipe cross-section area.

Interaction of centrifugal forces with the friction forces can be demonstrated by an example of a liquid motion in a pipe with its longitudinal axis curved by a circumference arch in horizontal flatness, under condition that the pipe cross-section is partially filled by the liquid. In the time of the liquid motion in such pipe the centrifugal forces press the liquid stream to external contour of the pipe. As a result a part of the pipe cross-section by its internal contour is found to be filled by air. At the same time because of the flatness symmetry of the curved pipe the tangential component of STHV forms the secondary pair-spiral flow of the liquid in the kind of two shrouds, moving symmetrically from sides of the liquid stream cross-section along the pipe cross-section contour to the pipe curvature centre and further forming the total flow along the symmetry flatness away from the pipe curvature centre. Core of these two liquid helices is filled by air.

Other example of the combined action of centrifugal forces and the friction forces is a rotation of liquid in the immovable cylindrical vessel with a bottom. This is the well-known problem on a rotation of tea in a cup, glass.

An elucidation of a character of the liquid motion in this case it is expediently to conduct on a basis of comparison of the above-mentioned motion with a rotation of the liquid jointly with a cylindrical vessel, possessing a bottom. In this second case the liquid is immovable relatively a lateral wall and a bottom of the vessel. Under action of centrifugal forces the liquid free surface takes a form of a paraboloid of rotation, i.e. this surface becomes concave. In the case of the liquid rotation in the immovable vessel the liquid also rises by the vessel lateral wall and one sinks by the rotation centre. At the same time the liquid height by the vessel wall is noticeable lesser; a form of the liquid free surface is close to the rotation paraboloid only in immediate proximity to the rotation axis: on the rest – greater – part the liquid free surface has a reverse curvature; this part of the liquid surface is formed by the spiral combs, directed from the vessel wall and aside of rotation and further to the liquid rotation axis.

For elucidating of a motion structure of the liquid, rotating in the immovable cylindrical vessel, the author has conducted a number of simple experiments by means of the glass cylindrical vessels, filled partially by water. For visualization of a flow by the liquid free surface as well as by a lateral wall and by a bottom of the vessel the author was putting in water a not great quantity of semolina and then he was tincturing the water with a few drops of Indian ink. At that the semolina drains, submerged preliminarily in water, were not colored by the ink and ones were good observed close by the vessel transparent walls. Rotation by hand of the liquid in the vessel and observation had allowed to ascertain the following:

- if the water depth not exceeds half radius of the vessel, then the water rotation leads to a formation of the secondary pair-spiral flow along the lateral wall and a bottom of the vessel similarly to the above-described flow in the flume, formed by two wall;

- if the water depth is commensurable with the vessel diameter, then the secondary pair-spiral flow is formed close to the vessel lateral wall completely; symmetry of a spreading of water by the vessel wall in both sides from horizontal flatness is similarly to the above-described spreading of the water free jet along the plate but in contrast to it the spirals of the secondary flow of water in the immovable cylindrical vessel are formed by the elementary vortex threads, as it is characteristically for the flume flow, described above; close by the vessel bottom and close by the water free surface the elementary vortex threads create the spiral flow aside the water rotation and away from the vessel lateral wall to the water rotation axis;

- if the water depth is much greater of the vessel diameter, then on the vessel lateral wall one can see the tracks of two or more the secondary pair-spiral streams; a flow close by the vessel bottom and close by the water free surface is already directed along a spiral to the water rotation axis;

- an increase of the rotation velocity of water increases a height of the water layer near the vessel lateral wall and at the same time it decreases the water depth close by the rotation axis right up to a uncovering of the vessel bottom; the water free surface takes a form close by a paraboloid of rotation under action of centrifugal forces; a line of a spreading of the secondary pair-spiral flow – only one – is displaced to a boundary of the vessel bottom with its lateral wall; the elementary vortex threads move in the lines of ascent along the vessel lateral wall away from the above mentioned boundary and then ones move along the spiral lines downward, forming the water free surface;

- the elementary vortex threads move in the same direction, corresponding to the water rotation, both to the vessel bottom and to the water free surface; the spiral motion of these threads in the indicated zones stipulates an increase of angular velocity of water close by its rotation axis in comparison with flow near the vessel lateral wall;

- the secondary pair-spiral flow is very steadily and one accompanies inertial rotation of water after a ceasing of the forced rotation.

Friction is a cause of formation of the secondary pair-spiral flows also by surface of cylindrical body, rotating in immovable liquid as it showed in fig. 121, 127 [7].

Prandtl in the above mentioned book [6] has also adduced the result of visualization of a flow in the straight flow element (pipe) with its cross-section in the kind of the rectangle, obtained by Nikuradze (1930).

In this case just as in the above described pipe of the triangle cross-section the total stream of the pair-spiral flows are directed into a vertex of every corner. Feature of the given case is in a lateral spreading of the liquid stream in the time of its forward motion along the large sides of such pipe cross-section. A flow in the corner zone can be presented as a flow in the flume, formed by two flatness under angle equaled to $90^0$. A flow along the flatness of two large sides of the rectangle can be presented as a flow in the flume with a flat bottom and flat lateral walls. As a result a sum of the flows in four two-flatness flumes, joined in pair by the flows in two flumes, possessed the bottom and lateral walls, reproduces theoretically the results of the above mentioned experiment.

The bath effect in the kind of the vertical crater, appearing in the time of an emptying of the bath through its bottom orifice is the well-known phenomenon.

For elucidating of this phenomenon it is sufficiently to imagine a liquid flow close by the bath bottom to its submerged orifice in the kind of several liquid jets, moving along the flat surface in radial direction to one centre (orifice). It is known (see section Solution of the given article) that a friction of the liquid free jet against the plate surface is accompanied by a spreading of this jet symmetrically in both sides from its longitudinal axis. Therefore radial motion of several such jets to the united discharge orifice must lead to a mutual restriction of its lateral spreading similarly to an action of lateral walls, restricting the lateral spreading of the liquid stream, moving in the flume. Such interaction of the adjacent jets leads, in one's turn, to a forming of the secondary ascending flow in the kind of radial combs along the contact line of the jets. A motion of these combs from the wall layer of the liquid intensive shear and inculcation of these combs in the liquid layer, being over the wall layer, leads to a conversion of these combs into the secondary pair-spiral flow along the boundaries of the adjacent jets. Under conditions of the blow-oscillating interaction of the adjacent jets between themselves the equilibrium of the symmetrical pair-spiral flow turns out unstable and one is accompanied by the irreversible transition to the secondary single-spiral flow, which one is displaced from a boundary of the adjacent jets to a surface of one of these jets. The same direction of the rotation velocity of the secondary single-spiral flow with the longitudinal motion velocity of the wall primary jet ensures not only stability of such secondary flow but also absorption of the weight flow and of the velocity head of the wall initial jet by this secondary flow.

This absorption is accompanied by a bending and by a displacing of longitudinal axis of the secondary flow aside to a beginning of the wall primary jet. Now the secondary single-spiral flow carries in itself all weight flow and velocity head of the wall flow and it moves near the bath bottom in a flat spiral to the discharging orifice, forming in it the vortex crater. Such flow is very stable and one can carry along in the rotary motion the upper layers of liquid. In that way, the bath effect can be formed by the real – viscous - fluid only.

 In a number of the here adduced examples it should be noted that Pitot device already 275 years allows to measure the STHV longitudinal component in a stream of the real – viscous – fluid, but not the total head, as it is presented in the modern hydromechanics and gas dynamics [6, 8]. Demonstration of the new approach to a mechanics of the real fluid motion can be unlimitedly continued. At the same time the author supposes that the adduced examples allow sufficiently clear to display the basic features of the stated solution as a basis of an overcoming of a problem on the secondary flows. Elucidation of a role of a friction and the new – vector – approach to the energy distribution in the real – viscous – fluid stream open the way to a physically adequate explanation of the causes of origin and development of the secondary flows and streams and its quantitative description. Side by side with the above mentioned articles [4, 5] - on a forming of a neck in cylindrical bar of plastic material in the time of its fracture under a tension and on the laminar stream instability - the developed conception on the stream total head vector ascertains a unity of the energy distribution in the problems of the solid body and fluid mechanics and mechanics of the contact interaction of the solid bodies between themselves and of the real fluid with the solid surface. Distinctive feature of the real fluid mechanics is bound with that the fluid possesses the freedom additional degrees in the kind of its internal mobility.

Thus the developed conception on a basis of the Physical Ensemble method in combination with the theses, stated in articles [9 – 17], allows for the first time to go over to a fruitful theoretical development of hydro-, aero-, and gas dynamics and to a creation of physically adequate apparatus for mathematical modeling of any phenomena, bound with the real fluid motion, using the modern mathematical apparatus and extensive material of the numerous experimental researches.

 Possibility of practical realization and a rightness of the new conception theses for a mathematical modeling of the motion dynamics of the viscous compressible fluid are corroborated by a creation of the first generation of fundamental version of the VeriGas computer program, worked out by the Physical-Technical Group specialists under the direct leadership of the author in September 2000.

**Final remarks**

Solution of a problem on the secondary flows, stated in the given article, quite correspond to I. Newton (320 years ago) instruction: "As a whole a difficulty of physics is in a recognizing of the Nature forces by the motion phenomena and then in an explaining of the rest of the phenomena by means of these forces".

The same solution excites also H. Pitot (275 years ago) question: "…why so simple and useful thought not came into a mind of the numerous scientists, which ones were producing experiments for research of the … flow … and were writing its scientific works to this subject".

**Acknowledgements**

The author dedicates the given article to Henri Pitot memory, who is one of the outstanding scientists, engineers and inventors.

---

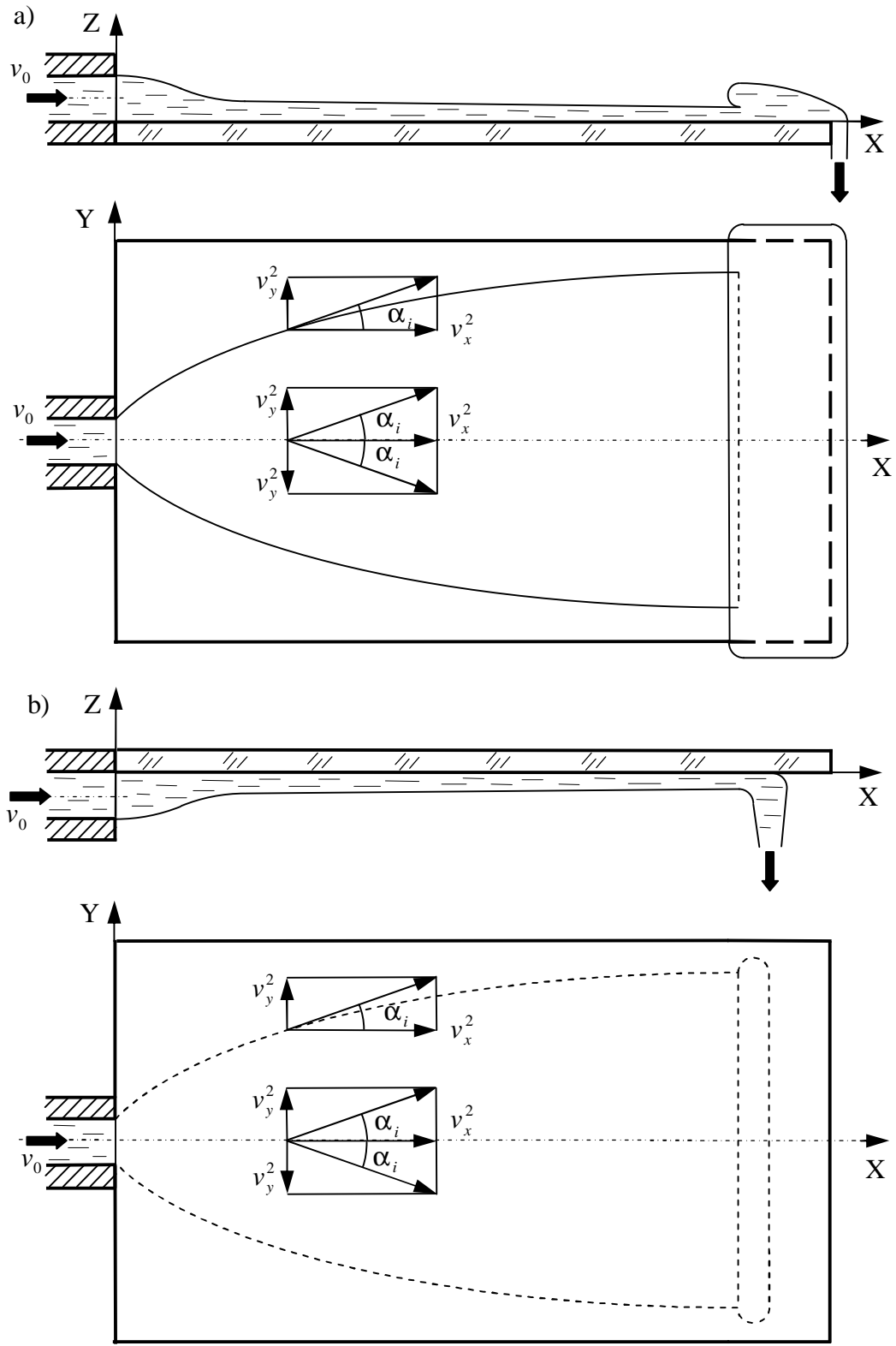

The diagrams of a motion of the liquid free jet along the upper (a) and lower (b) surfaces of the horizontal plate